\documentstyle[aps,preprint,tighten,epsf]{revtex}

\def\fns{{F_n^{(s)}}}
\def\ffkks{{F_{KK^*}^{(s)}}}
\def\sst#1{{\scriptscriptstyle #1}}
\def\beq{\begin{equation}}
\def\eeq{\end{equation}}
\def\beqa{\begin{eqnarray}}
\def\eeqa{\end{eqnarray}}
\def\MeV{\nobreak\,\mbox{MeV}}
\def\GeV{\nobreak\,\mbox{GeV}}

\def\mn{{m_{\sst{N}}}}
\def\mns{{m^2_{\sst{N}}}}
\def\mks{{m_{\sst{K}}^2}}

\def\bra#1{{\langle#1\vert}}
\def\ket#1{{\vert#1\rangle}}
\def\pbar{{\bar{p}}}

\newcommand{\nomi}{\hphantom{-}}
\def\FOS{{F_1^{(s)}}}
\def\FTS{{F_2^{(s)}}}
\newcommand{\dida}[1]{/ \!\!\! #1}

\def\PRC#1{{{\em   Phys. Rev.} {\bf C#1} }}
\def\PRD#1{{{\em   Phys. Rev.} {\bf D#1} }}
\def\PRL#1{{{\em   Phys. Rev. Lett.} {\bf #1} }}
\def\NPA#1{{{\em   Nucl. Phys.} {\bf A#1} }}
\def\NPB#1{{{\em   Nucl. Phys.} {\bf B#1} }}

\def\PLB#1{{{\em   Phys. Lett.} {\bf B#1} }}

\def\ZPC#1{{{\em   Z. f\"ur Phys.} {\bf C#1} }}

\begin{document}
\draft
\preprint{INT \# DOE/ER/40561-345-INT97-00-184}

\title{\bf $K^*$ Mesons and Nucleon Strangeness }
\vskip 5 cm
\author { L.L. Barz$^{1,2}$, H. Forkel$^{3,4}$, H.-W. Hammer$^{5,6}$,
F.S. Navarra$^1$, M. Nielsen$^1$, and  M.J.
Ramsey-Musolf$^{6,7}$\footnote{National Science Foundation Young 
Investigator}\\
[0.5cm]
{\it $^1$Instituto de F\'{\i}sica, Universidade de S\~{a}o Paulo}\\
{\it C.P. 66318,  05315-970 S\~{a}o Paulo, SP, Brazil} \\[0.1cm]
{\it $^2$Faculdade de Engenharia de Joinville, Universidade Estadual
de Santa Catarina}\\
{\it 82223-100 Joinville, SC, Brazil} \\[0.1cm]
{\it$^3$European Centre for Theoretical Studies in Nuclear Physics
and Related Areas},\\
{\it Villa Tambosi, Strada delle Tabarelle 286, I-38050 Villazzano,
Italy} \\[0.1cm]
{\it$^4$Institut f{\"u}r Theoretische Physik, Universit{\"a}t
Heidelberg},\\
{\it Philosophenweg 19, D-69120 Heidelberg, Germany} \\[0.1cm]
{\it $^5$ TRIUMF, 4004 Wesbrook Mall, Vancouver, B.C.,
Canada V6T 2A3} \\[0.1cm]
{\it $^6$ Institute for Nuclear Theory, University of Washington,
Seattle, WA 98195, USA} \\[0.1cm] 
{\it $^7$ Department of Physics, University of Connecticut, Storrs, 
CT 08629 USA}
}

\maketitle
\vspace{3cm}
\begin{abstract}
We study contributions to the nucleon strange quark vector
current form factors from intermediate states containing
$K^{*}$ mesons. We show how these contributions may be comparable
in magnitude to those made by $K$ mesons, using methods complementary
to those employed in quark model studies. We also analyze the degree
of theoretical uncertainty associated with $K^{*}$ contributions.
\\

\end{abstract}

\vspace{1cm}
\newpage
\section{Introduction}
\label{intro}

The role played by virtual $q\bar{q}$ pairs in the low-energy
structure of hadrons remains one of the outstanding questions
for hadron structure physics. Despite the evidence for important
$q\bar{q}$ sea effects obtained with deep inelastic scattering,
the experimental manifestations of explicit sea-quark effects
at low energies are minimal. Partial explanations for this absence
have been given using a non-relativistic quark model framework
by the authors of Ref. \cite{Gei90}, who noted that in
the adiabatic approximation, virtual $q\bar{q}$ pairs renormalize
the string tension and, therefore, do not have any discernable
impact on the low-lying spectrum of hadronic states. Similarly,
virtual $q\bar{q}$ effects -- in the guise of virtual mesonic
loops -- which could conceivably lead to large $\rho-\omega$ and
$\phi-\omega$ mixing were shown to cancel at second order in
strong couplings when a sum is performed over a tower of virtual
hadronic states \cite{Gei91}. The latter result provides insight into the
applicability of the OZI Rule to $V$-$V'$ mixing despite the na\"\i ve
scale of $q\bar{q}$ effects expected at one-loop order.

Nevertheless, several mysteries involving $q\bar{q}$ pairs remain to
be solved. Of particular interest are those involving nucleon matrix
elements of strange quark operators, $\bra{N}\bar{s}\Gamma s\ket{N}$.
The latter explicitly probe properties of the $q\bar{q}$ sea at low energies,
since the nucleon contains no valence strange quarks. Moreover, the
mass scale associated with $s\bar{s}$ pairs -- $m_s\sim\Lambda_{QCD}$ --
implies that such pairs live for sufficiently long times and propagate
over sufficiently large distances to produce observable effects when
probed explicitly. In this respect, $s\bar{s}$ pairs stand in contrast
with, {\em e.g.}, $c\bar{c}$ pairs, whose effects one expects to be
suppressed by powers of $\Lambda_{QCD}/m_c\sim 0.1$ \cite{Kap88}.

Some support for these simple-minded expectations is provided by
determinations of $\bra{N}\bar{s} s\ket{N}$ from the $\pi N$ \lq\lq
sigma" term \cite{Che76} and of $\bra{N}\bar{s}\gamma_\mu\gamma_5 s\ket{N}$
from polarized deep inelastic scattering \cite{EMC} and neutrino-nucleus
quasi-elastic scattering \cite{Ahr87}. The former suggests that roughly
15\% of the nucleon mass is generated by $s\bar{s}$ pairs, while the
latter implies that strange quarks contribute about 30\% of the total
quark contribution to the nucleon spin\footnote{Theoretical
uncertainties associated with SU(3) breaking qualify the conclusions
drawn from deep inelastic scattering experiments, however.}. Measurements
of $\bra{N}\bar{s}\gamma_\mu s\ket{N}$, which would provide information
about the strange quark contribution to the nucleon magnetic moment and
rms radius are presently underway at MIT-Bates \cite{MIT},
Mainz \cite{Mai}, and the Jefferson Laboratory \cite{TJN}. The first
results for the strangeness magnetic form factor have been reported in
Ref. \cite{MIT}. One expects this set of $\bra{N}\bar{s}\Gamma
s\ket{N}$ determinations to provide a clearer picture of the $q\bar{q}$
sea than obtained from existing spectroscopic data alone.

Despite over a decade of theoretical efforts to study nucleon strangeness,
the theoretical understanding of s-quark matrix elements remains in its
infancy. In the case of $\bra{N}\bar{s}\gamma_\mu s\ket{N}$, a plethora
of predictions have been reported in the literature
\cite{Lat,Mod,Jaf89,Had,Gei97,Mrm97a,Mus97b,Mrm97c,HWH97}.
While a few lattice results have been obtained by different groups\cite{Lat}, 
they are not entirely consistent with each other nor with the recent first
results for the \lq\lq strange magnetic moment" obtained by the SAMPLE
collaboration\cite{MIT}. The
remaining predictions -- based generally on QCD-inspired
nucleon models \cite{Mod,Gei97} or low-energy truncations of QCD in a hadronic
basis  \cite{Jaf89,Had,Mrm97a}
-- display a broad range in magnitude and sign. Recently, it has
been shown why such truncations -- either in the strong coupling constant 
($g$) expansion (loop order) \cite{Mus97b,Mrm97c}
or hadronic excitation energy ($\Delta E$) \cite{Gei97} -- are
untrustworthy and may produce misleading results. The implication of these
studies is that the intuitively appealing picture of a kaon cloud around
the nucleon does not suffice to describe $s\bar{s}$ fluctuations in
the nucleon. It appears that
one must include both the full set of virtual hadronic
intermediate states \cite{Gei97} as well as the full set of higher-order
rescattering effects for a given state
\cite{Mus97b,Mrm97c} in order to obtain a physically
realistic prediction. In principle, corrections to the leading order
truncations in $\Delta E$ and $g$ could be accounted for by the appropriate
low-energy constants in chiral perturbation theory (CHPT); however, chiral
symmetry does not afford a determination of the low-energy constant
relevant to nucleon vector current strangeness \cite{Mrm97a}. Hence, one must
understand in some detail the short-distance strong interaction mechanisms
responsible for the low-energy structure of the strange quark sea.

In the present study, we amplify on the themes of
Refs. \cite{Gei97,Mrm97a,Mus97b,Mrm97c} by
studying the $K^{\ast}$ contribution to $\bra{N}\bar{s}\gamma_\mu s\ket{N}$.
Our objective is two-fold: (i) to illustrate, using an alternative
framework to that of Ref. \cite{Gei97},
how inclusion of higher-lying intermediate
states may alter conclusions obtained when only the lightest \lq\lq
OZI-allowed" fluctuation is included, and (ii) to demonstrate the
theoretical uncertainty associated with computing higher-lying contributions.
For these purposes, we restrict ourselves to second order in the strong
meson-baryon coupling, $g$, when treating hadronic amplitudes $N\to
YK^*$ {\em etc.}, fully cognizant of the shortcomings such a truncation
entails. In fact, the kind of analysis of higher-order effects reported
in Ref. \cite{Mrm97c} for the $K\bar{K}$ intermediate state does not appear
feasible at present for higher lying states. Consequently, some form of
model-dependent truncation is necessary when treating these states, and
we do not, therefore, pretend to make any reliable numerical predictions.
Rather, we use the ${\cal O}(g^2)$
(one-loop) truncation to illustrate the two main points
stated above. In this respect, our study is similar in spirit to that of
Ref. \cite{HWH97}, where a comparison at one-loop order was made to show that
contributions from intermediate states containing no valence strangeness
($3\pi$) and those containing valence s-quarks ($K\bar{K}$) may be comparable
in magnitude.

In order to estimate the degree of theoretical uncertainty one has in
the numerical prediction for the $K^{\ast}$ contribution, we use two
approaches to carry out the calculation: (a) an explicit one-loop
calculation, where form factors are included at hadronic vertices and
the intermediate state $\bar{s}\gamma_\mu s$ matrix elements are taken
to be point-like, and (b) a computation using dispersion relations, in
which the $N\bar{N}\to KK,\ KK^{\ast},\ K^{\ast}K^{\ast}$ amplitudes
are computed in the Born approximation but form factors are included
at the $\bar{s}\gamma_\mu s$ insertions. These computations are outlined,
respectively, in Sections II and III. In Section IV, we discuss the
results of the calculations and compare with the conclusions drawn in
Ref. \cite{Gei97}.

\section{One-loop calculation}
\label{ext}

The first \lq\lq kaon cloud" estimates of $\bra{N}\bar{s}\gamma_\mu s\ket{N}$
were obtained from the amplitudes associated with the diagrams of Fig. 1,
where only the contributions for $B=B'=\Lambda, \Sigma$ and $M=M'=K$  were
included \cite{Had}. Here we consider the next heaviest contributions by
including the
octet of spin-one mesons as well as the pseudoscalars, and compute the
following amplitudes where, in each case, $B=B'=\Lambda$ or $\Sigma$: 
(1a) for $M=K^{\ast}$; (1b) for $M=M'=K^{\ast}$; (1b) for
$M=K$, $M'=K^{\ast}$; (1c) for $M=K^{\ast}$. As we discuss below, the 
diagrams (1c) are required for consistency with the Ward-Takahashi identities.

	The resulting contributions to the strange-quark vector current
matrix element are embodied in the Dirac and Pauli form factors
defined via

\beq
\langle N(p^\prime) | \bar s\gamma_\mu s| N(p) \rangle =
{\bar U}(p^\prime) \left[F_1^{(s)}(q^2)
\gamma_\mu + i{\sigma_{\mu\nu}q^\nu\over
2m_N}F_2^{(s)}(q^2)\right]U(p)\ \ \ \ ,
\label{ma}
\eeq
where $U(p)$ denotes the nucleon spinor. Recall
that $F_1^{(s)}(0)=0$, due to the zero strangeness charge of
the nucleon. The leading nonvanishing moments of the corresponding Sachs
form factors
\beqa
G_E^{(s)}(q^2)= F_1^{(s)}(q^2)+ {q^2\over4m_N^2}F_2^{(s)} (q^2),
\\
G_M^{(s)}(q^2)= F_1^{(s)}(q^2)+F_2^{(s)}(q^2)\;
\eeqa
are the strangeness radius
\beq
\langle r_s^2 \rangle_S = \left.6{d\over dq^2}
G_E^{(s)}(q^2)\right|_{q^2=0} \; ,
\label{rs}
\eeq
and the strangeness magnetic moment
\beq
\mu_s=G_M^{(s)}(0)=F_2^{(s)}(0)\;  .
\label{mus}
\eeq
For future reference, we note that the Sachs radius $\langle r_s^2 \rangle_S$
is related to the corresponding Dirac radius as
\beq
\label{sachsdirac}
\langle r^2_s \rangle_S= \langle r^2_s \rangle_\sst{D} +
\frac{3}{2\mns} \mu_s\,.
\eeq

In order to extend the $K - \Lambda$ loop framework to
include $K^*$-meson contributions, we start from the
meson baryon effective lagrangians
\begin{eqnarray}
\label{1aa}
{\cal L}_{MB} & = &-ig_{ps} \bar{B} \gamma_5 B K\ \ \ , \\
{\cal L}_{VB} & = & -g_v(\bar{B} \gamma_\alpha B V^\alpha +
{\kappa\over 2m_N}\bar{B}
\sigma_{\alpha\beta} B \partial^\alpha V^\beta)\; ,
\label{la}
\end{eqnarray}
where $B$, $K$, and $V^\alpha$ are the baryon, kaon,  and $K^*$
vector-meson fields respectively, $m_N=939\MeV$ is the
nucleon mass and $\kappa$ is the ratio of  tensor to vector
coupling, $\kappa=g_t/g_v=3.26$,  with $g_v/\sqrt{4\pi}=
 -1.588$ \cite{hol89}. The strength of the pseudoscalar
coupling is $g_{ps}/\sqrt{4\pi} = -3.944$ \cite{hol89}.

In order to account in some way for the finite extent of the hadrons
appearing in the loops of Fig. 1, we include form factors at the
hadronic vertices. For simplicity, we adopt a monopole form
\beq
F(k^2)={m^2-\Lambda^2\over k^2-\Lambda^2} \ \ \ .
\label{ff}
\eeq
Although there is no rigorous justification for this choice,
form factors of this type for the $KN\Lambda$ and
$K^* N \Lambda$ vertices are used in the
Bonn potential. Their cut-off parameters are determined from
hyperon-nucleon scattering data \cite{hol89}:
$\Lambda_{K^*}=2.2$ (2.1), $\Lambda_K=1.2 (1.4) \GeV$ with masses $m_K= 495$ 
MeV and $m_{K^*}=895\MeV$\cite{PDG}. 
The numbers in parenthesis denote
values obtained in an alternate model for the baryon-baryon
interaction.The momentum of the $K^*$ is $k$. These form factors render
all the following loop integrals finite and reproduce the
on-shell values of the mesonic couplings (since $F(m^2)=1$) .

In the presence of electroweak fields the non-local
meson-baryon interaction of Eqs. (\ref{1aa}-\ref{ff})
gives rise to vertex currents. In order
to maintain gauge invariance we introduce the photon field
by minimal substitution of the momentum variable in the form
factors\footnote{As noted in \cite{ohta,Mrm97a}
and elsewhere this procedure is not
unique since the Ward-Takahashi identity does not restrict the
transverse part of the vertex.}. This procedure generates the
nonlocal seagull vertex \cite{ohta,wan96,Mrm97a}
\beq
i\Gamma_{\mu\alpha}^{(s)}(k,q)=ig_vQ_{K^*} (q\pm 2k)_\mu
{F(k^2)-F((q\pm k)^2) \over(q\pm k)^2-k^2}\left[\pm\gamma_\alpha
+{i \kappa\over{2m_N}}
\sigma_{\alpha\beta}k^\beta\right] \; ,
\label{seag}
\eeq
where the upper/lower signs correspond to an incoming/outgoing
vector meson (with index $\alpha$), $Q_{K^*}=-1$ is the $K^*$
strangeness charge, and $q$ is the photon momentum.

Due to the derivative in eq. (\ref{la}), the minimal substitution
also generates an additional seagull vertex (even in the absence of
meson-nucleon form factors)
\beq
i\Gamma_{\mu\alpha}^{(v)}(k)=\pm{g_v \kappa Q_{K^*} \over 2 m_N}
F((q\pm k)^2) \sigma_{\alpha\mu}\; ,
\label{ver}
\eeq
where the sign convention is the same as above.

The diagonal matrix elements of $\bar{s}\gamma_\mu s$ for strange
mesons and baryons is straightforwardly determined by current
conservation and the net strangeness charge of each hadron.
The structure of the s-quark current spin-flip transition from $K$ to $K^*$
is
\beq
\langle K^*_a(k_1,\varepsilon)|{\overline s}\gamma_\mu s|K_b(k_2)\rangle=
{\ffkks(q^2)\over m_{K^*}}\epsilon_{\mu\nu\alpha\beta}k_1^\nu
k_2^\alpha \varepsilon^{*\beta}\delta_{ab}\; ,
\eeq
where $a$ and $b$ are isospin indices, $\varepsilon^\beta$ is
the polarization vector of the $K^*$, and $k_1, k_2$ are the
meson momenta. In a loop calculation, $\ffkks(q^2)$ is
taken to be a constant equal to its value at the photon point.
In order to estimate this constant, we follow Ref.~\cite{gm} and
assume $\ffkks(q^2)$ to be dominated at low-$q^2$ by the
lightest $I^G(J^{PC})=0^-(1^{--})$ vector mesons\footnote{The validity
of this assumption is discussed in more detail in the following section.}:
\beq
\label{omegaphi}
{\ffkks(q^2)\over m_{K^*}}=-\sum_{V=\omega,\phi}
{G_{K^*VK}S_V\over q^2-m_V^2}
\; ,
\eeq
where $G_{K^*VK}$ are the couplings of the vector meson
$V$ to $K$ and $K^*$. $S_V$ determines the strength of the
strange-current conversion into $V$:
\beq
\langle0|{\overline s}\gamma_\mu s|V\rangle=S_V
\varepsilon_\mu
={m_V^2\over f_V}{f_V\over f_V^{(s)}}\varepsilon_\mu\; .
\eeq

>From the known isoscalar electromagnetic couplings
$f_{\omega,\phi}$ one can delineate the corresponding
strange-current couplings with the help of a simple quark
counting prescription based on flavor symmetry
\cite{Jaf89}:
\beq
\label{fvrel}
{f_\omega\over f_\omega^{(s)}}=-
\sqrt{6}{\sin\epsilon\over\sin{(\theta_0+
\epsilon)}}\;\;\;\;\;\; ,\;\;\;\;\;\;\;{f_\phi\over f_\phi^{(s)}}
= -\sqrt{6}{\cos\epsilon\over\cos{(\theta_0+\epsilon)}}\; ,
\eeq
Here $\epsilon=0.053$ \cite{jain} is the mixing angle between
the pure $\overline{u}u+ \overline{d}d$ and $\overline{s}s$
states and the physical vector mesons $\omega$ and $\phi$,
and $\theta_0$ is the ``magic angle'' defined by $\sin^2\theta_0
=1/3$. From the above we find $f_\omega / f_\omega^{(s)}
 = -0.21$ and $f_\phi/ f_\phi^{(s)} = -3.11$. Combined with
the strong couplings $G_{K^*\phi K} = -8.94\GeV^{-1}$ and
$G_{K^*\omega K} = 6.84 \GeV^{-1}$ estimated in Ref. \cite{gm}
we finally obtain
\beqa
\ffkks(0)=1.84.
\eeqa

After these preparations\footnote{Note also that the small SU(3) values for
the $\Sigma K N$ couplings \cite{hol89} lead to a strong
suppression of the contributions from $\Sigma K$ intermediate
states \cite{Had}. This argument does not affect,
however, the $\Sigma^* K$ and $\Sigma^* K^*$ contributions.},
we can evaluate the $K^*$ loop contributions to the nucleon's
strangeness radius and magnetic moment. Explicit expressions for
the loop amplitudes are given in Appendix A. The results for the
different diagrams are listed in Table I. The implications of
these results are discussed in Section IV.

\section{Dispersion relation calculation}
\label{dispa}

An alternative approach to computing virtual hadronic contributions
to strange quark form factors is the use of dispersion relations
(DR's). In principle, DR's provide a method for including information
beyond second order in $g$, both via the strong amplitudes
$N\to YK^{*}\to N$ and through the form factors $F_n^{(s)}$
describing the intermediate state matrix elements $\bra{Y}\bar{s}
\gamma_\mu s\ket{Y}$, $\bra{K^{*}}\bar{s}\gamma_\mu s\ket{K}$,
$\ldots$. The one-loop calculation of Section II is equivalent to
the use of a DR in which the strong amplitudes $N\to YK^{*}\to N$
are computed in the Born approximation and the form factors assumed
to be point-like: $F_n^{(s)}(q^2)=F_n^{(s)}(0)=
{\hbox{const}}$\footnote{The equivalence holds only when the hadronic
form factors of Eq. (\ref{ff}) are set to unity.}.

The inclusion of rescattering and resonance effects in the
$N\to YK^{*}\to N$ amplitude would require the existence of sufficient
data for $KN\to NK\pi,\ldots$ or $N\bar{N}\to KK\pi, KK\pi\pi$ {\em etc.}
to permit analytic continuation of these amplitudes to the unphysical
regime as needed for the dispersion relation. Although such a program
is feasible to some degree for the $K\bar{K}$ intermediate state
\cite{Mrm97c}, it does
not appear practical at present for the case of higher mass strange
mesons of interest here. Consequently, we include 
the amplitude for $N\to YK^{*}\to N$
at the level of the Born approximation. In the case of the $F_n^{(s)}(q^2)$,
however, it is possible to introduce some structure beyond the point-like
approximation, albeit in a model-dependent way. Our strategy for doing
so is discussed below.

First, we review the formalism for treating strangeness form factors with
DR's. We write an unsubtracted dispersion relation for the
Pauli form factor $F^{(s)}_2$ and subtract the one for the Dirac
form factor $F^{(s)}_1$ once at $t=0$ (where $F^{(s)}_1$ vanishes,
see above):
\begin{eqnarray}
  \label{disp1}
F^{(s)}_1(t) &=& \frac{t}{\pi}\int\limits_{t_0}^\infty dt'
\frac{\hbox{Im}\ F^{(s)}_1(t')}{t'(t' -t)}\ \ \ , \\
F^{(s)}_2(t) &=& \frac{1}{\pi}\int\limits_{t_0}^\infty dt'
\frac{\hbox{Im}\ F^{(s)}_2(t')}{t' -t} \ \ \ ,
\nonumber
\end{eqnarray}
where $t\equiv q^2$.
The cut along the real $t$-axis starts at the threshold $t_0$
of a given multi-particle intermediate state, as {\em e.g.} $t_0 =4\mks$
for the $K\bar{K}$ state. From Eqs.  (\ref{disp1}) one expects
that contributions from the lightest intermediate states will
mainly determine the behavior of the form factors at $t=0$.
The imaginary part of the form factors is readily obtained
by means of a spectral decomposition.
Since the matrix elements $\bra{N(p)}\bar{s}\gamma_\mu s \ket{N(p')}$
and $\bra{N(p);\bar{N}(\pbar)}\bar{s}\gamma_\mu s\ket{0}$
are simply related by crossing symmetry, we write the spectral
decomposition for the latter one as \cite{Mrm97c},
\begin{eqnarray}
\label{spec_t}
& &{\hbox{Im}}\ \bra{N(p);\bar{N}(\pbar)}\bar{s}\gamma_\mu s\ket{0} =
{\hbox{Im}}\ \bar{U}(p)\left[F_1^{(s)}(t)\gamma_\mu + i{\sigma_{\mu\nu}(p+
\pbar)^\nu  \over 2\mn} F_2^{(s)}(t)\right]V(\pbar)\\
& &\qquad\rightarrow
{\pi\over\sqrt{Z}}(2\pi)^{3/2}{\cal N}\sum_{n}
\bra{N(p)}\bar{J_N}(0)\ket{n}\bra{n}\bar{s}\gamma_\mu s\ket{0} V(\pbar)
\delta^4(p+\pbar-p_n)\, ,
\nonumber
\end{eqnarray}
where ${\cal N}$ is a spinor normalization factor,
$Z$ is the nucleon's wave function renormalization constant,
and $J_N(x)$ is a nucleon source.
Nonzero contributions arise only from physical states
$\ket{n}$ with the same
quantum numbers as the current $\bar{s}\gamma_\mu s$,
{\em i.e.} $I^G(J^{PC})=0^-(1^{--})$ and zero baryon number. These
asymptotic states $\ket{n}$ in the above sum do not
explicitly contain resonances. Resonance contributions arise
via the matrix elements $\bra{N(p)}\bar{J_N}(0)\ket{n}$ and
$\bra{n}\bar{s}\gamma_\mu s\ket{0}$.
In the vector meson dominance
approximation, one assumes the product of the two matrix elements
in Eq. (\ref{spec_t}) to be strongly peaked near vector meson masses.
This approximation has been used in several pole analyses of the
strange vector form factors \cite{Jaf89}.

The lightest contributing intermediate states are purely mesonic:
$3\pi$, $5\pi$, $7\pi$, $K\bar{K}$, $K\bar{K}\pi$, $9\pi$,
$K\bar{K}\pi\pi$, $\ldots$, in order. Intermediate baryon states
$N\bar{N}$, $\Lambda\bar{\Lambda},\ldots$
appear with significantly higher thresholds, $t_0$.
In the present study,
we restrict ourselves to the strange states and consider corrections
to the $K\bar{K}$ state. The first such corrections (in order of
threshold) are those involving the $K\bar{K}\pi$ and $K\bar{K}\pi\pi$
intermediate states. In the previous section, these states were
included using the narrow resonance approximation: $K\bar{K}\pi
\to K^{*}\bar{K}$ and $K\bar{K}\pi\pi\to K^{*}\bar{K}^{*}$. In
order to make contact with the loop results of Section II as well
as with the calculation of Ref. \cite{Gei97} where 
in effect the same approximation
was made, we adopt the narrow resonance approximation here.  We also
include the $\Lambda\bar{\Lambda}$ and $\Sigma\bar{\Sigma}$ intermediate
states, even though they are not among the lightest in the series, in
order to compare the DR results with those of the loop and quark
model calculations, which contain these states.

As noted earlier,
we also include the strong amplitudes $\bra{N}\bar{J}_N(0)\ket{n}$
at the level of the Born approximation. For the matrix elements
$\bra{n}\bar{s}\gamma_\mu s\ket{0}$, parameterized by form factors
$F_n^{(s)}(t)$, we go beyond the point-like approximation,
\beq
F_n^{(s)}(t)\equiv F_n^{(s)}(0) \equiv F_n^0\ ,
\eeq
of the one-loop and quark model calculations
by allowing for some structure in the form factors. For the mesonic
intermediate states, we make a simple vector meson dominance (VMD)
{\em ansatz}.  This {\em ansatz} is well justified for the $K\bar{K}$ state,
following from
$e^+e^-\to K\bar{K}$ cross section data \cite{Del81} and simple flavor
rotation arguments \cite{Jaf89}.
The $e^+ e^-\to K\bar{K}$ data indicates a strong peak in the
vicinity of the $\phi$ resonance, with a subsequent rapid fall-off
as $q^2$ (time-like) increases away from $m_\phi^2$.
Inclusion of a VMD-type form factor peaked near the $\phi$-resonance
significantly affects the $K\bar{K}$ component of the spectral functions
[Eqs. (\ref{spec_t})] and
the resulting contribution to the strangeness moments as
compared with the use of a point-like form factor.

In the case of the $KK\pi\sim
KK^{*}$ and $KK\pi\pi\sim K^{*}K^{*}$
states, we take the $\fns(t)$ to be dominated by
either the $\phi(1020)$ or the $\phi'(1680)$. Following Ref. \cite{Mus97b}
we write
\begin{equation}
\label{fk_vdm}
| \fns(t)_{VDM}| =F_n^0
\left\{ {(\xi^2)^2+M^2\Gamma^2\over [(\xi^2-t)^2
+M^2\Gamma^2]}\right\}^{1/2}\,, \label{phidom}
\end{equation}
where $M=m_{\phi}=1020$ MeV or $m_{\phi'}=1680\pm 20$ MeV,
$\Gamma= \Gamma_{\phi}=4.43\pm 0.05$ MeV or $\Gamma_{\phi'}= 150\pm 50$ MeV
are the total widths of the $\phi$ or $\phi'$ \cite{PDG},
and $\xi^2\equiv M^2-\Gamma^2 /4$.
As we note below, we need only the magnitude of the form factor in
the present calculation, as the $n\to N\bar{N}$ amplitudes are real
in the Born approximation. Because the states $KK\pi\sim
KK^{*}$ and $KK\pi\pi\sim K^{*}K^{*}$ contribute to
the DR of Eq. (\ref{disp1}) for $t_0> m_\phi$, we expect higher mass
vector mesons to play a significant role in
the $\fns(t)$ in the region of integration.

The case for  $\phi'$ dominance is most convincing
for the $K\bar{K}\pi$ intermediate state. Data for
$\sigma(e^+e^-\to K_S^0 K^\pm \pi^\mp)$ in the range $1.4 \leq \sqrt{s}
\leq 2.18$ GeV display a pronounced peak near $\sqrt{s}=1.680$ GeV
\cite{Man82}.
Furthermore, Dalitz plot analyses imply that the final state is dominated
by a $K^{\ast}K\leftrightarrow KK\pi$ resonance. The OZI rule implies that
the $\phi'$ is nearly a pure $s\bar{s}$ state, while SU(3) relations
and data for $\sigma(e^+e^-\to \rho\pi;\sqrt{s}\approx 1.65)$ constrain
the $\omega'-\phi'$ mixing angle to deviate by less than $10^\circ$ from
ideal mixing \cite{Buo82}.
While the tails of the $\rho(770)$, $\omega(780)$, and
$\phi(1020)$ affect details of the peak structure, the dominant effect
is that of the $\phi'$ \cite{Buo82}.
In the absence of any other structure in $\sigma(e^+e^-\to KK\pi)$
in the region $t>t_0$, we conclude that $\ffkks(t)$ should also
be dominated by the $\phi'(1680)$. Indeed, the $\omega\phi$ model of
Eq. (\ref{omegaphi}), which is credible for low-$t$, is inconsistent
with annihilation data for $t> t_0$. Using it in this region would
generate an artificial suppression of the $KK\pi$ spectral function.

With these considerations in mind, it is straightforward to determine
the normalization $F_{KK^*}^0$ appearing in Eq. (\ref{fk_vdm}). Following
the notation of Ref. \cite{gm}, we obtain
\beq
\label{fkks_zero}
F_{KK^*}^0=  G_{KK^*\phi'} m_{K^*}/f_{\phi'}^{(s)}\ \ \ ,
\eeq
where $1/f_{\phi'}^{(s)}\approx -3/f_{\phi'}$, and
$G_{KK^*\phi'}$ is the strong $\phi'\to KK^*$ coupling. The latter
may be obtained from $\Gamma(\phi'\to KK^*)$ which, for a single final
charge state is \cite{gm}
\beq
\Gamma(\phi'\to KK^*) = {|G_{KK^*\phi'}|^2\over 12\pi} |k_F|^3\ \ \ ,
\eeq
where $k_F=463$ MeV is the $K$ or $K^*$ CM momentum. Assuming
$\Gamma(\phi'\to {\hbox{all}})$ is dominated by $\Gamma(\phi'\to KK^*)$
\cite{PDG}, we obtain $|G_{KK^*\phi'}|\approx 3.8$ GeV$^{-1}$.

Similarly, the $\phi'$ electronic width determines $f_{\phi'}$:
\beq
\Gamma(\phi'\to e^+e^-) = {4\pi\over 3} \alpha^2 {M_{\phi'}\over
f_{\phi'}^2} \ \ \ .
\eeq
Analyses of $e^+e^-$ data yield $\Gamma(\phi'\to e^+e^-)=0.7$ keV
\cite{Buo82,Bis91},
from which we obtain $f_{\phi'}\approx 23$. The factor of $-3$ appearing
in the relation between $f_{\phi'}$ and $f_{\phi'}^{(s)}$ assumes
ideal mixing [see Eq. (\ref{fvrel})]. Allowing for a small deviation
$|\epsilon|< 10^\circ$ does not change our results appreciably,
especially since the $\omega'$ is not observed to decay to $KK\pi$.

Substituting these results into Eq. (\ref{fkks_zero}) yields
$F_{KK^*}^0=\ffkks(0)=0.43$, to be compared with the value
$\ffkks(0)=1.84$ used in loop calculation. We emphasize the latter
value results from assuming only the $\omega$ and $\phi$ contribute
to $\ffkks(t)$, whereas the former is obtained when {\em only} the
$\phi'$ is included. Depending on the relative
phase of the $(\rho\omega\phi)$ and $(\rho\omega\phi)'$ contributions
in $e^+e^-\to KK\pi$, the $\phi'$ will either increase or decrease the
point-like value (1.84) for this form factor by about 25\% . At the $KK^*$
threshold, the $\phi'$ contribution to $\ffkks$ is about half as large
as that from the $\phi$, but becomes nearly five times larger in
the vicinity of $t=m_{\phi'}^2$. For purposes of estimating the
$t$-dependence of $\ffkks$ in the region $t>t_0$, then, inclusion
of only the $\phi'$ appears to be a reasonable
approximation\footnote{A more sophisticated treatment, including the
tails of the $\phi$ and $\omega$, would -- as in the purely EM
case -- affect the shape of the form factor near the $\phi'$ peak
and the resultant $KK^*$ spectral function.}. We note in passing
that our estimate of the $\phi'$ contribution carries an uncertainty
of 25\% or more, as the experimental values for $\Gamma(\phi'\to KK\pi,
e^+e^-)$ carry experimental errors of $\geq 25\% $ \cite{Cle94}.

The implications of $e^+e^-$ data for $F_{K^*}^{(s)}(t)$ are less
clear. To our knowledge, there exists no annihilation data giving
$K^*K^*$ branching ratios. In $e^+e^-\to KK\pi\pi$ ($1.4\leq
\sqrt{s}\leq 2.18$), for example, the $K\pi$ invariant mass
distribution is consistent with production of only one $K^*$ per
event \cite{Cor82}. Consequently, the data cannot be used to infer a $K^*$
EM or strangeness form factor for $t>t_0$, and we must rely on a model.
Given the evidence for $\phi'$ dominance of $\ffkks(t)$ and for
$\phi$ dominance of $F_K^{(s)}(t)$ as well as the
absence of experimental observation of any
$0^-(1^{--})$ $s\bar{s}$ mesons with mass $\geq 2 m_{K^*}$,
it is natural to assume that the $t$-dependence of $F_{K^*}^{(s)}(t)$
is governed by the tails of the known $s\bar{s}$ vector mesons.
For simplicity, we include only one $s\bar{s}$ resonance -- either
the $\phi$ or the $\phi'$ -- using the form of Eq. (\ref{fk_vdm}).
The normalization $F_{K^*}^0=|Q_{K^*}|$. In the DR
results displayed in Table I, we quote a range of values, the limits
of which correspond to using either the $\phi$ or $\phi'$.
A more realistic parameterization of $F_{K^*}^{(s)}(t)$ is likely to
include some linear combination of $\phi$ and $\phi'$ poles, as well
as small contributions from the $\omega$ and $\omega'$.
Existing information does not permit us to determine this linear
combination. Consequently, we use the ranges appearing in Table I
to estimate the uncertainty in the $K^*K^*$ contribution
associated with lack of knowledge of the $K^*$ strangeness form factor.

For the intermediate hyperon form factors, we are aware of no electromagnetic
data to provide guidance for the choice of $F_n^{(s)}(t)$. We therefore
work in analogy with the proton EM form factors, since both $F_B^{(s)}(t)$
($B=\Lambda,\ \Sigma$) and $F_\sst{PROTON}^\sst{EM}(t)$ involve matrix elements
of vector currents having unit conserved charge in the states of interest.
Consequently, we adopt the standard dipole form factor for the
Dirac strangeness form factors of the intermediate hyperons. Since the
corresponding strange magnetic couplings are unknown, we omit magnetic
form factors altogether. Because the resulting contributions to the
strangeness moments are generally small compared to the mesonic
contributions, we do not expect the uncertainty associated with $F_B^{(s)}(t)$
to be problematic.

Under these assumptions, our calculation proceeds as follows.
The spectral functions entering Eqs. (\ref{disp1}) have the
general form
\beq
\label{genform}
{\hbox{Im}}\  F(t) = {\hbox{Re}}\ \left[A_{J=1}^n(t)\ \fns(t)^*\right]=
	|A_{J=1}^n(t)| \ |\fns(t)| \ (1+\gamma_n)\ ,
\eeq
where $A_{J=1}^n$ is the appropriate combination of $J=1$ partial waves
for the process $n\to N\bar{N}$ and $\gamma_n$ is a correction arising from
the difference in phases between the amplitude $A_{J=1}^n$ and
$\fns$ \cite{Mus97b}.
This correction can vary between $-2$ and 0 and depends on
$t$. At present, we are unable to determine $\gamma_n$ for the
intermediate states considered here, and set $\gamma_n=0$ to obtain
an upper bound.

To compute the $A_{J=1}^n(t)$ in Born approximation,
we calculate the imaginary
parts of the diagrams (a) and (b) in Fig. 1 assuming point-like
strangeness form factors, $\fns(t)\equiv 1$.
We neglect the hyperon-nucleon mass difference and
take $m_\sst{Y}=\mn$.
The seagull diagrams do not have an imaginary part, so we obtain
no contributions from diagrams 1c.
Furthermore, from Eq. (\ref{spec_t}) the individual contributions
are manifestly gauge invariant in this approach.
We calculate the imaginary parts of the corresponding diagrams
with cutting rules \cite{cutru} and insert them into the dispersion
relations Eqs. (\ref{disp1}). To obtain the imaginary parts it is
convenient to consider the crossed $t$-channel matrix element
$\bra{N(p);\bar{N}(\pbar)}\bar{s} \gamma_\mu s\ket{0}$. The
generic form of such a diagram is shown in Fig. 2.
The different choices for the internal lines I, II,
and III are shown in Table II.
The equivalent of the previous kaon loop result is recovered
if the internal lines are chosen as in case 1 and 2. In the
following, we outline our calculation for the cases 3 - 5.  In
case 3 and 4, both kaons have been replaced by $K^*$ vector
mesons, while one kaon and one $K^*$ contribute in case 5.

We choose to work in the center-of-momentum (CM) frame of the
nucleon-antinucleon pair, where $q=(\omega,\vec{0})$.
The loop diagrams
lead to a physical reaction for $t \geq 4 \mns$, which is
the minimal energy required for the creation of a $\bar{N}N$-pair,
and we have $p'=(\omega/2,\vec{p'})$
and $p=(\omega/2,-\vec{p'})$ with $p_t=|\vec{p'}|=\sqrt{t/4-\mns}$.
We define the contribution of a particular Feynman diagram with
vertex function $\Gamma_\mu$ as
\begin{equation}
  \label{vert}
{\cal M}^{(i)}_\mu= -i\, \bar{u}(p') \Gamma^{(i)}_\mu v(p)\,.
\end{equation}
These vertex functions are then multiplied by the strangeness form
factor $|F^{(s)}(t)_{VDM}|$ from above as indicated by Eq. (\ref{genform}).
Our choice for the momenta of the internal lines is indicated
in Fig. 2.

For the cases 3 - 5 we obtain the vertex functions
shown in Appendix B.
The imaginary part of $\Gamma_\mu^{(i)}$ is always finite;
hence, the divergencies of the $d^4k$ integrals are without
consequences. The vertex functions $\Gamma^\mu$ have
branch cuts on the real axis for $ t \geq (m_I+m_{II})^2$.
Their real part is continuous, such that the discontinuity associated
with the cut is reflected only in the imaginary part.
In the CM-frame of the nucleon and antinucleon, we have to
calculate
\begin{equation}
\label{discon}
{\hbox{Im}}\,\Gamma^\mu = \frac{1}{2\,i}\Delta\Gamma^\mu =
\frac{1}{2\,i}\lim_{\delta \to 0}
\left(\Gamma^\mu(\omega+i\delta)-\Gamma^\mu(\omega-
i\delta)\right) \,.
\end{equation}
In particular, we obtain the discontinuity $\Delta\Gamma^\mu$
using the Cutkosky rules \cite{cutru} by
cutting the lines I and II, i.e. by replacing the propagators of these
lines by $\delta$ functions,
\begin{equation}
\label{cuma}
\frac{1}{p^2 - m^2 + i\varepsilon} \longrightarrow
 -2\,\pi\,i\,\theta(p_0)\, \delta(p^2 - m^2)\; .
\end{equation}
As a consequence, the discontinuity arises when the particles
I and II in Fig. 2 are on-shell.
Due to the delta functions, the $d^4 k$ integration
covers only a finite part of the $k$ space, leading to a finite
value of the integral. Next we write
$d^4k$ as $dk_0\, k^2 dk \,d\Omega_k$ and use the delta
functions to carry out the $dk_0$ and $dk$ integrations.
Moreover, the $d\Omega_k$ integration involves only
$x$, the cosine of the angle between $\vec{k}$ and $\vec{p'}$.
The denominator of the remaining propagator acquires the
structure $z + x$, where $z$ depends on the particles internal to
the loop.
\beqa
{\rm Case} \; 3\quad &:& \qquad z = \frac{2(2\,\mns-m_\sst{K^*}^2)
 -t}{4p_t^2}=-(1+
\frac{m_\sst{K^*}^2}{2p_t^2})\,\\
{\rm Case} \; 4 \quad &:& \qquad z= \frac{2 m_\sst{K^*}^2 -t}{4p_t
q_t}\\
& &\qquad q_t = \sqrt{t/4 -m_\sst{K^*}^2}\nonumber\\
\mbox{Case} \; 5\quad &:& \qquad z = \frac{\mks +m_\sst{K^*}^2-2q_t^2
-2\sqrt{(q_t^2 +\mks)(q_t^2+m_\sst{K^*}^2)}-t}{8 p_t q_t} \\
& & \qquad q_t = \frac{1}{2\sqrt{t}}\sqrt{t^2 +(m_\sst{K^*}^2- \mks)^2
-2t(m_\sst{K^*}^2+ \mks)} \nonumber
\eeqa
Finally, ${\hbox{Im}}\,\Gamma_\mu$ can be expressed through Legendre
functions of the second kind, and, using the relation
\begin{equation}
  \label{f_zer}
{\hbox{Im}} \,\Gamma_\mu = \gamma_\mu {\hbox{Im}}\,F_1 +i
\frac{\sigma_{\mu\nu}}{2m} q^\nu {\hbox{Im}}\, F_2 \; ,
\end{equation}
the contributions to the imaginary parts of the Dirac and Pauli
form factors for $t \geq 4\,\mns$, respectively, can be
identified. The emerging spectral functions
are valid for $t \geq 4\,\mns$.
The dispersion integrals, however, start at $t_0=(m_I + m_{II})^2$,
with $m_I$
and $m_{II}$ the masses of the loop particles I and II, respectively.
Consequently, the imaginary parts of the diagrams with two internal
meson lines have to be analytically continued into
the unphysical region $(m_I + m_{II})^2 \leq t < 4\,\mns$,
by replacing the momentum
$p_t = \sqrt{t/4 -\mns}$ by $i\,p_{-} = i\sqrt{\mns -t/4}$.
Similarly, the variables $z$ become complex ($z \to i \xi$),
and the Legendre functions of the second kind must be analytically
continued as well.

Inserting now the imaginary parts and their
analytical continuations in the unphysical region
into the dispersion relations of
Eq. (\ref{disp1}), we obtain the $KK^*$ and $K^*K^*$ contributions
to the strangeness form factors of the
nucleon. In particular, the dispersion relations for the $K^*$ loop
contributions to the strangeness radius and magnetic moment
read
\begin{eqnarray}
\label{rhosi}
\langle r^2_s \rangle_D &=&{6\over\pi}\int_{t_0}^\infty dt {{\hbox{Im}}\,
	\FOS(t)\over t^2} \\
\label{musi}
\mu^{(s)}&=& {1\over\pi} \int_{t_0}^\infty dt {{\hbox{Im}}\, \FTS(t)\over t}\,,
\end{eqnarray}
where $\langle r^2_s\rangle_D$ is related to
the Sachs radius via Eq. (\ref{sachsdirac}).
For most of the intermediate states
considered here, the dispersion integrals in Eqs. (\ref{rhosi}, \ref{musi})
converge when a non-pointlike form for the $F_n^{(s)}(t)$ is employed.
However, the tensor $K^*NB$ ($B=\Lambda, \Sigma$) coupling renders the
$K^{*}K^*$ divergent even when the VDM form factor is included. To
regulate this integral, we note that the unitarity of the S-matrix implies
that the $N\bar{N}\to K^{*}K^{*}$ amplitude is bounded in magnitude
for scattering in the physical region, $t> 4m_N^2$. The Born approximation
for this amplitude does not respect this boundedness property, signalling
the importance of higher-order rescattering
corrections \cite{Mus97b}. At present,
since we wish only to obtain an estimate for the $K^{*}$ contributions,
we replace the $A_{J=1}^n(t>4 m_N^2)$ by its value at
the physical threshold, $A_{J=1}^n(t=4 m_N^2)$. We make the same replacement in
the integrals for the $KK^{*}$ intermediate state. This procedure
leads to a crude upper bound on the contribution to the integrals from
the integration region $t> 4 m_N^2$.  

The results of the DR estimates of the various contributions are
quoted in Table I. The DR results for the $K\bar{K}$ contribution given
in Table I were obtained using the rigorous unitarity bound.
We stress that the $K^*$ results give rough upper bounds on the various
contributions, not only because of the boundedness of the
strong amplitudes but also because the phase difference correction,
$\gamma_n$, is not known.  We also do not compute the total contributions from
the various states, as we cannot presently determine their relative
phases. Only in the one-loop calculation of the previous section
are the relative phases fixed by the model.

\section{ Discussion and Conclusions}
\label{disc}

The results shown in Table I illustrate the two primary conclusions of
our analysis: (i) contributions from higher mass intermediate states
to the strangeness moments are not necessarily small compared with those
from the lightest ``OZI allowed" state $K\bar{K}$ ; (ii) estimating
these higher mass contributions can entail a significant degree of
theoretical uncertainty.

In the one-loop model,
the $K^*$ contributions can be as much as an order of
magnitude larger than those from the kaon loop. The origin of
this result can be traced to two factors: the tensor coupling
of the $N\Lambda K^*$ vertex is much larger than the $N
\Lambda K$ coupling, and the cut-off of the Bonn form
factor involving the $K^*$ is about twice as large as that
involving the kaon ($\Lambda_K = 1.2$ GeV).
In the case of the former, omitting
the tensor coupling reduces the contribution to the strangeness
radius by a factor of five to ten and yields a near exact
cancellation between the $KK$, $KK^{*}$, and $K^{*}K^{*}$
contributions. In the case of the strange magnetic moment,
the large $K^{*}K^{*}$ and $KK^{*}$ contributions drop by two orders of
magnitude when $\kappa$ is set to zero.

The effect of the larger cut-off is particularly emphasized
in graphs which contain
derivative ({\em i.e.} tensor) couplings of the $K^*$. These couplings
bring in additional powers of the loop momentum $k$ and the
corresponding loop integrals therefore receive larger
contributions from $k$ of the order of the cut-off.  However,
the importance of loop momenta above $\sim$2 GeV points to
weaknesses of the one loop approximation. As we discuss in
more detail below, the large $K^*K\Lambda$ and $K^*K^*\Lambda$
contributions (1b) appear to result from un-physical, un-realistically
large values of the integrand for large loop momenta. Physically
realistic contributions from these intermediate states are likely
to be much smaller.

In fact, the DR contributions from the $KK^*$ and $K^*K^*$ states
are significantly smaller in magnitude than those generated in the
loop model, though they are still comparable to, or larger than, the
$K\bar{K}$ contribution. The reduction in the magnitude of these
contributions from the loop model estimate reflects two factors:
the boundedness of the $n\to N\bar{N}$ scattering amplitude in the
physical region and the presence of more realistic, non-pointlike
$\fns(t)$. Although we have only implemented the boundedness crudely
for the $KK^{*}$ and $K^{*}K^{*}$ states,
the requirement that the partial waves are bounded in the physical
region ($t>4\mns$) is a rigorous one, following from the unitarity of
the S-matrix. Since a one-loop calculation is equivalent to a DR
in which the $\fns(t)$ are taken to be pointlike and the $A_{J=1}^n$
computed in the Born approximation, the one-loop results do not
respect the boundedness requirement. The presence of hadronic form
factors [Eqs. (\ref{ff})] does not remedy this violation since they
preserve the on-shell form for the $n\to N\bar{N}$ amplitudes.

In the $K\bar{K}$ case, the unitarity violation of the one-loop
calculation was shown to be a serious one \cite{Mus97b}. For the
intermediate states containing a $K^*$, this violation appears
to be all the more serious, as a comparison of the DR and loop
results suggests. The tensor coupling of the $K^*$ to baryons
weights the $K^*K\to N\bar{N}$ and $K^*K^*\to N\bar{N}$ amplitudes
more strongly in the physical region, relative to the un-physical
region ($t_0\leq t\leq 4\mns$), than in the $K\bar{K}\to N\bar{N}$ case.
Consequently, the physical region contributes a substantial fraction
of the entries (1b) for the $K^*K$ and $K^*K^*$ states (80\% of the total
in the $K^*K^*$ case) -- even after the imposition of a crude bound
on the $A_{J=1}^n$ and inclusion of non-pointlike $F_K^{(s)}(t)$.
Had we not imposed even our rough bound, the $K^*K^*$ contribution
to $\langle r_s^2\rangle_D$, for example, would have been five times
larger. We conclude
that the large contributions to the strangeness moments resulting from
the one-loop model are not physically realistic.

We emphasize that the DR calculation given here -- though containing
more physical information than the one-loop model -- remains incomplete.
A rigorous unitarity bound for the $K^*K$ and $K^*K^*$ amplitudes
remains to be implemented, as has been done in the $K\bar{K}$ case. 
More importantly,
the impact of higher order (in $g$) rescattering corrections and possible
resonance effects in the $A_{J=1}^n(t_0\leq t\leq 4\mns)$ must also be
estimated. In the $K\bar{K}$ case, these effects significantly
enhance the $\langle r_s^2\rangle$
contribution over the entry $KK\Lambda$ (1b) in
Table I \cite{Mrm97c}. This enhancement arises primarily from a near threshold
$\phi(1020)$-resonance in the $K\bar{K}\to N\bar{N}$ amplitude. Similarly, we
expect inclusion of $K^*K$ and $K^*K^*$ rescattering and $\phi'$ resonance
effects in the $A_{J=1}^n$ to modify the $K^*K$ and $K^*K^*$ entries
in Table I. Unfortunately, sufficient $KK\pi\to N\bar{N}$ (or $KN\to
K\pi N$) and $KK\pi\pi\to N\bar{N}$ ($KN\to KN\pi\pi$ {\em etc.}) data
do not presently exist to afford a model-independent
determination of these effects.

Given that higher mass contributions to the strangeness moments need not
be small compared to that from the $K\bar{K}$, it is desireable to reduce
the theoretical uncertainty in the former as much as possible.
The $K^*K^*\Lambda$ (1b) entry hints at the level of this uncertainty.
Our \lq\lq reasonable range" for this contribution allows for about a factor
of four to seven variation, 
which follows from the choice of different, but reasonable,
$K^*$ strangeness form factors. Based on our previous study of the $K\bar{K}$
contribution, as well as the behavior of the scattering amplitudes in
the physical region,
we may reasonably expect a similar level of uncertainty
associated with the presently unknown rescattering and resonance effects
in the $A_{J=1}^{K^*K,\ K^*K^*}$.

To summarize, we have estimated $K^*K$ and $K^*K^*$ contributions to the
nucleon strangeness moments, using two approaches which complement the
quark model calculation of Ref. \cite{Gei97}. Our results confirm the
conclusions
reached in that work that higher mass hadronic states can be as important
as the $K\bar{K}$ state and that a calculation of the strangeness moments
based on a truncation in $\Delta E$ is not reliable. Similarly, we illustrate
the significant theoretical ambiguities involved in estimating these higher
mass contributions -- particularly those associated with effects going
beyond ${\cal O}(g^2)$ and with the intermediate state strangeness form
factors. In this study, we have taken the first steps toward including
the latter in a realistic way. We find that inclusion of physically reasonable
parameterizations of the $\fns(t)$ can appreciably affect the $K^*K$ and
$K^*K^*$ contributions. Even here, however, our efforts are limited by
a lack of existing EM data. In the case of higher-order and resonance
effects in the strong amplitudes, it should be evident that simple models
which do not account for them can produce physically unrealistic estimates
of the higher mass intermediate state contributions. Clearly, more
sophisticated approaches are needed in order to understand how $s\bar{s}$
pairs live as virtual hadronic states.

\acknowledgements

We would like to thank D. Drechsel and N. Isgur for useful discussions.
This work has been supported in part by FAPESP and CNPq. M.N.
would like to thank the Institute for Nuclear Theory at the
University of Washington for its hospitality and H.F.
acknowledges an HCM grant from the European Union and
a DFG habilitation fellowship. MJR-M has been supported
in part under U.S. Department of Energy contracts \#
DE-FG06-90ER40561 and \# DE-AC05-84ER40150 and under a
National Science Foundation Young Investigator Award. HWH
has been supported by the Deutsche Forschungsgemeinschaft
(SFB 201) and the German Academic Exchange Service
(Doktorandenstipendium HSP III/ AUFE).

\vspace{0.5cm}

\appendix
\section{Vertex Functions: Loops}

In the following appendices we list  the explicit expressions
for the one-loop diagrams considered in Section \ref{ext}.

They are numbered as in the figures: 
 (1a) for $M=K^{\ast}$ and $B=B'=\Lambda,\Sigma$;
(1b) for $M=M'=K^{\ast}$ and $B=B'=\Lambda,\Sigma$; (1b) for
$M=K$, $M'=K^{\ast}$, and $B=B'=\Lambda,\Sigma$; (1c) for $M=K^{\ast}$
and $B=B'=\Lambda,\Sigma$. 

\begin{eqnarray}
\Gamma^{(1a)}_\mu(p^\prime,p)& =& ig^2_v Q_B \int \frac{d^4k}
{(2\pi)^4}  (F(k^2))^2 D^{\alpha\beta}(k)\left(\gamma_\alpha 
+i{\kappa\over2m_N}
\sigma_{\alpha\nu}k^\nu\right)
S(p^\prime-k) \gamma_\mu\times 
\nonumber\\*[7.2pt]
&&S(p-k) \left(\gamma_\beta-i{\kappa\over2m_N}\sigma_{\beta\gamma}k^\gamma
\right) \; ,
\label{1a}
\end{eqnarray}
\begin{eqnarray}
\Gamma^{(1b)}_\mu(p^\prime,p)& =&- ig^2_v Q_{K^*}  \int 
\frac{d^4k}{(2\pi)^4} F((k+q)^2)F(k^2)  D^{\alpha\lambda}(k+q)
D^{\sigma\beta}(k)\left(\gamma_\alpha + \right.
\nonumber\\*[7.2pt]
&+&\left.i{\kappa\over2m_N}
\sigma_{\alpha\nu}(k+q)^\nu\right)
[(2k+q)_\mu \, g_{\sigma\lambda}-(k+q)_\sigma g_{\lambda\mu}-k_\lambda 
g_{\sigma\mu}]\times
\nonumber\\*[7.2pt]
&&S(p-k)\left(\gamma_\beta-i{\kappa\over2m_N}\sigma_{\beta\gamma}
k^\gamma\right)   \; ,\;{\mbox{for $M=M^\prime=K^*$}}
\nonumber\\*[7.2pt]
& =&-{g_vg_{ps}F_{K^*K}^{(s)}(0)\over m_{K^*}}
\epsilon_{\mu\nu\lambda\alpha}\int \frac{d^4k}{(2\pi)^4}\left\{F((k+q)^2)F_K
(k^2)D^{\alpha\beta}(k+q)\times\right.
\nonumber\\*[7.2pt]
&&\Delta(k^2)(k+q)^\nu k^\lambda \left(\gamma_\beta +i{\kappa\over2m_N}
\sigma_{\beta\delta}(k+q)^\delta\right)S(p-k)\gamma_5+
\nonumber\\*[7.2pt]
&+&F(k^2)F_K((k+q)^2)D^{\alpha\beta}(k)\Delta((k+q)^2)k^\nu 
(k+q)^\lambda \gamma_5 \times
\nonumber\\*[7.2pt]
&&\left.S(p-k)\left(\gamma_\beta -i{\kappa\over2m_N}\sigma_{\beta\delta}
k^\delta\right)\right\}\; ,\;{\mbox{for $M=K\;,M^\prime=K^*$}}
\label{1b}
\end{eqnarray}
\begin{eqnarray}
\Gamma^{(1c)}_\mu(p^\prime,p)& =& g^2_v Q_{K^*}  \int \frac{d^4k}
{(2\pi)^4} F(k^2) D^{\alpha\beta}(k) \left\{i
\left[\frac{ (q+2k)_\mu}{ (q+k)^2-k^2} 
\left(F(k^2)\, - F((k+q)^2)\right) \times \right.\right.
\nonumber\\*[7.2pt]
& & \left(\gamma_\alpha +i{\kappa\over2m_N}\sigma_{\alpha\nu}k^\nu\right)
S(p-k)\left(\gamma_\beta-i{\kappa\over2m_N}\sigma_{\beta\gamma}k^\gamma\right) 
- \frac{ (q-2k)_\mu}{ (q-k)^2-k^2} (F(k^2)+
\nonumber\\*[7.2pt]
&&\left.-F((k-q)^2))\left(\gamma_\alpha +i{\kappa\over2m_N}\sigma_{\alpha\nu}
k^\nu\right)
S(p^\prime-k)\left(\gamma_\beta-i{\kappa\over2m_N}\sigma_{\beta\gamma}k^\gamma
\right) \right] \; +
\nonumber\\*[7.2pt]
&+&\;{\kappa\over2m_N}\left[F((k+q)^2)\sigma_{\alpha\mu} 
S(p-k)\left(\gamma_\beta-i{\kappa\over2m_N}\sigma_{\beta\gamma}k^\gamma
\right)\right. + 
\nonumber\\*[7.2pt]
&-&\left.\left.F((k-q)^2)\left(\gamma_\alpha +i{\kappa\over2m_N}
\sigma_{\alpha\nu}k^\nu
\right)S(p^\prime-k)\sigma_{\beta\mu}\right]\right\} \; ,
\label{1c}
\end{eqnarray}

\noindent
In the above equations we define $p^\prime=p+q$ and use the
notation $D_{\alpha\beta}(k)=(-g_{\alpha\beta}  +
k_\alpha k_\beta/m_{K^*}^2)(k^2-m_{K^*}^2+i\epsilon)^{-1}$
for the
$K^*$ propagator, $\Delta(k^2)=(k^2-m_K^2+i\epsilon)^{-1}$
for the kaon propagator, $S(p-k) = (p\kern-.5em\slash- k\kern-
.5em\slash-m_B+ i\epsilon)^{-1}$ for the hyperon, $B$,
propagator with mass  $m_\Lambda=1116\MeV$, $m_\Sigma=1193\MeV$ and 
strangeness charge $Q_B =1$.

\section{Vertex Functions: Dispersion Calculation}

Here, we display the vertex functions for the dispersion
relation calculation of Section III. We require
the product of propagator denominators and $|F^{(s)}(t)_{VDM}|$
for the cases 3-5.  This product is abbreviated by
\beqa
\label{denab}
{\cal D}_3 &=&\left\{[(k-q/2)^2 -\mns-i\epsilon][(k+q/2)^2 -\mns-
i\epsilon] \right.\\ & & \left.[(p'-k-q/2)^2 -m_\sst{K^*}^2-i\epsilon]
\right\}^{-1}\,|F^{(s)}(t)_{VDM}|\,,\nonumber
\eeqa
for case 3 and accordingly for cases 4 and 5. The vertex
functions are labelled as in section III (Table I). We obtain:
\begin{itemize}
\item[]Case 3 ($K^*K^*B$\ 1a) :
\begin{eqnarray}
\Gamma_\mu^{(3)} &=& -iQ_B g_v^2\int\frac{d^4 k}{(2\pi)^4}\,
(\gamma_\alpha +\frac{i \kappa }{2\mn}\sigma_{\alpha\nu}
(p'-k-q/2)^\nu)\\ &\nomi&(\dida{k}+\dida{q}/2+\mn)
\gamma_\mu(\dida{k}-\dida{q}/2+\mn) \nonumber\\
&\nomi&(\gamma_{\alpha'}-\frac{i \kappa}{2\mn}
\sigma_{\alpha'\nu'}(p'-k-q/2)^{\nu'}) \nonumber \\ &\nomi&
(g^{\alpha\alpha'}-(p'-k-q/2)^\alpha
(p'-k-q/2)^{\alpha'}/m_\sst{K^\ast}^2)\, {\cal D}_3\; \nonumber
\end{eqnarray}
\item[]Case 4 ($K^*K^*B$\ 1b) :
\begin{eqnarray}
\Gamma_\mu^{(4)} &=& -iQ_{K^*} g_v^2 \int\frac{d^4 k}{(2\pi)^4}\,
(\gamma_{\beta'}+\frac{i \kappa}{2\mn}\sigma_{\beta'\nu}
(k+q/2)^\nu)\\
&\nomi& (g^{\beta'\beta}-(k+q/2)^{\beta'}(k+q/2)^{\beta}/
m_\sst{K^*}^2) \nonumber \\ &\nomi&
(g^{\alpha\alpha'}-(k-q/2)^\alpha (k-q/2)^{\alpha'}/m_\sst{K^*}^2)
\,(\dida{p'}-\dida{k}-\dida{q}/2+\mn)
\nonumber \\ &\nomi&(2k_\mu g_{\beta\alpha}
-g_{\beta\mu}(k+q/2)_\alpha-g_{\alpha\mu}(k-q/2)_\beta)
\nonumber\\
&\nomi&(\gamma_{\alpha'}-\frac{i \kappa}{2\mn}\sigma_{\alpha'\nu'}
(k-q/2)^{\nu'}){\cal D}_4\; \nonumber
\end{eqnarray}
\item[]Case 5 ($KK^*B$\ 1b) :
\begin{eqnarray}
\Gamma_\mu^{(5)} &=&-2 g_{ps}g_v \frac{F_{K^*K}^{(s)}(0)}{m_\sst{K^*}}
\int\frac{d^4 k}{(2\pi)^4}\,
(\gamma_{\beta'}+\frac{i\kappa}{2\mn}\sigma_{\beta'\nu}
(k+q/2)^\nu)\\
& & (g^{\beta'\beta}-(k+q/2)^{\beta'}(k+q/2)^{\beta}/m_\sst{K^*}^2)
\nonumber \\ & &\epsilon_{\sigma\beta\rho\mu}
(k+q/2)^\sigma q^\rho (\dida{p'}-\dida{k}-\dida{q}/2+\mn) \gamma_5
{\cal D}_5\; \nonumber
\end{eqnarray}
\end{itemize}

\newpage

\begin{table}[hb]
\begin{center}
\begin{tabular}{|c||c|c|c|c|} \hline
Contribution & $\langle r_s^2 \rangle_D \; (\mbox{fm}^2) \;   \; $
loop  &  $|\langle r_s^2 \rangle_D |\;  (\mbox{fm}^2)   \;   \; $
DR & $\mu_s \; $ loop & $|\mu_s|  \; $ DR \\
\hline \hline
$KKB$\ 1a &0.006  & $0.001$ &-0.107  & $0.023$ \\
$KKB$\ 1b &-0.009  & $0.036$ &-0.078  & $0.143$ \\
$KKB$\ 1c &-0.004  & 0 &-0.069 & 0 \\    
$KKB$\ tot & $-0.007$ &  & $-0.24$ &  \\
 &  &  &  &  \\
$K^*K^*B$\ 1a & $ 0.075$  & 0.001 & $-2.283$  & 0.053  \\
$K^*K^*B$\ 1b & $-0.038$  & $0.003\to 0.012$ & $-2.343$  & $0.059\to 0.408$\\
$K^*K^*B$\ 1c & $-0.007$ & $0$ & 0.499 & $0$ \\
$K^*K^*B$\ tot & $0.030$ &  & $-4.127$ &  \\
 &   &  &  &  \\
$KK^*B$\ 1b & 0.078 & 0.035  & 1.015 & 0.425  \\
 &  &  &  &  \\
total  & $0.101$ &  & $-3.352$ &  \\
\hline
\end{tabular}
\end{center}
\caption{\label{kstartab4} Intermediate state contributions
to the strange magnetic
moment $\mu_s$ and the electric strangeness radius $\langle
r_s^2 \rangle_D $. The contributions are labelled according
to the diagrams in Fig. 1 and the intermediate state particles.}
\end{table}

\begin{table}[htb]
\begin{center}
\begin{tabular}{|c||c|c|c|} \hline
Case  & I & II & III \\ \hline \hline
1 & $\qquad K\qquad$ & $\qquad K\qquad$ & $\qquad\Lambda,\;\Sigma\qquad$ \\
2 & $\Lambda,\;\Sigma$ & $\Lambda,\;\Sigma$ & $K$ \\
3 & $\Lambda,\;\Sigma$ & $\Lambda,\;\Sigma$ & $K^*$ \\
4 & $K^*$ & $K^*$ & $\Lambda,\;\Sigma$ \\
5 & $K$ & $K^*$ & $\Lambda,\;\Sigma$ \\
\hline
\end{tabular}
\end{center}
\caption{\label{kstartab1} Particles assigned to the internal lines in
the loop diagram of Fig. 2.}
\end{table}

\newpage

\begin{figure}
\begin{center}
\epsfysize=14cm
~\epsffile{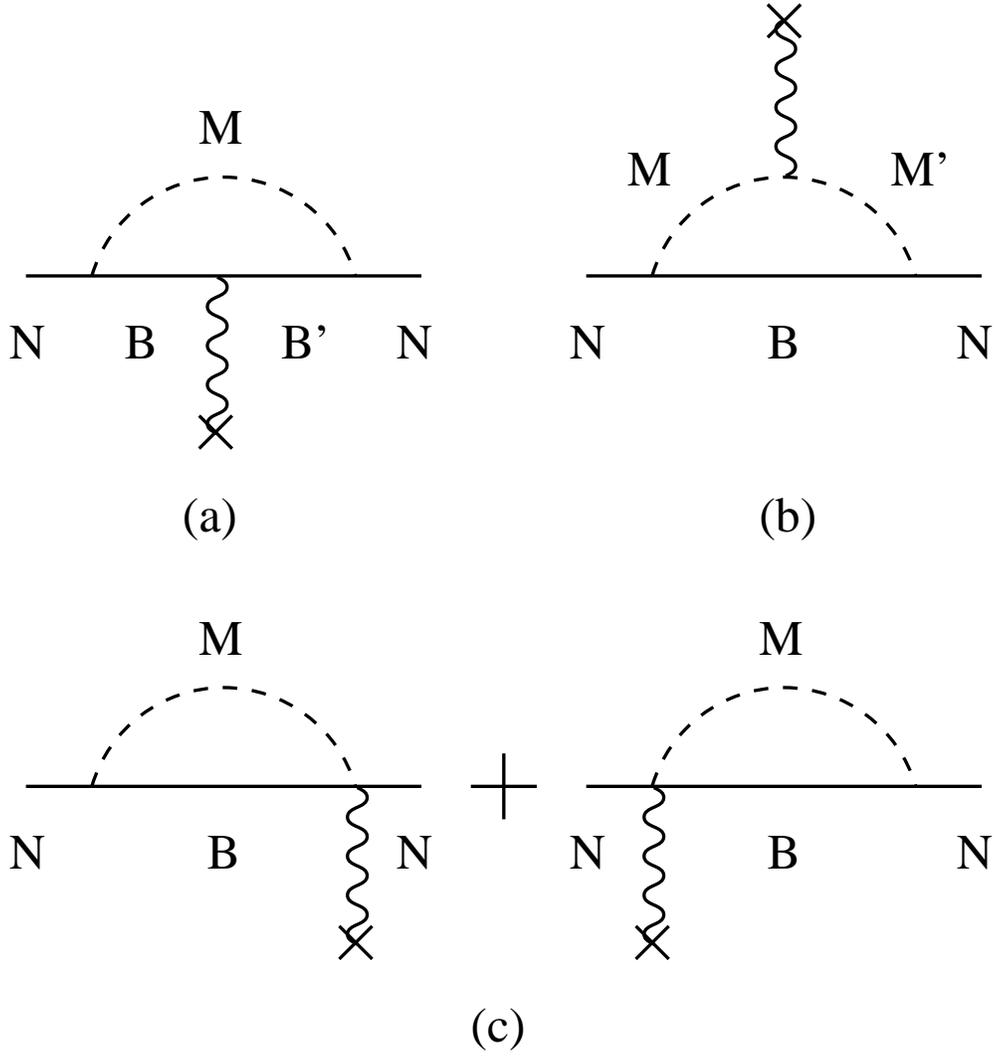}
\end{center}
\caption{Diagrams representing $KB$ and $K^*B$ contributions
to nucleon strangeness form factors. Solid line represents
baryon (B) and dashed line indicates meson (M, M'). Fig. (1a) gives
$B\bar{B}$ contribution; (1b) gives $K\bar{K}$, $K^*K$, and
$K^*K^*$ contributions; (1c) indicates seagull contributions
required by gauge invariance.}
\label{fig-1}
\end{figure}

\begin{figure}
\begin{center}
\epsfysize=10cm
~\epsffile{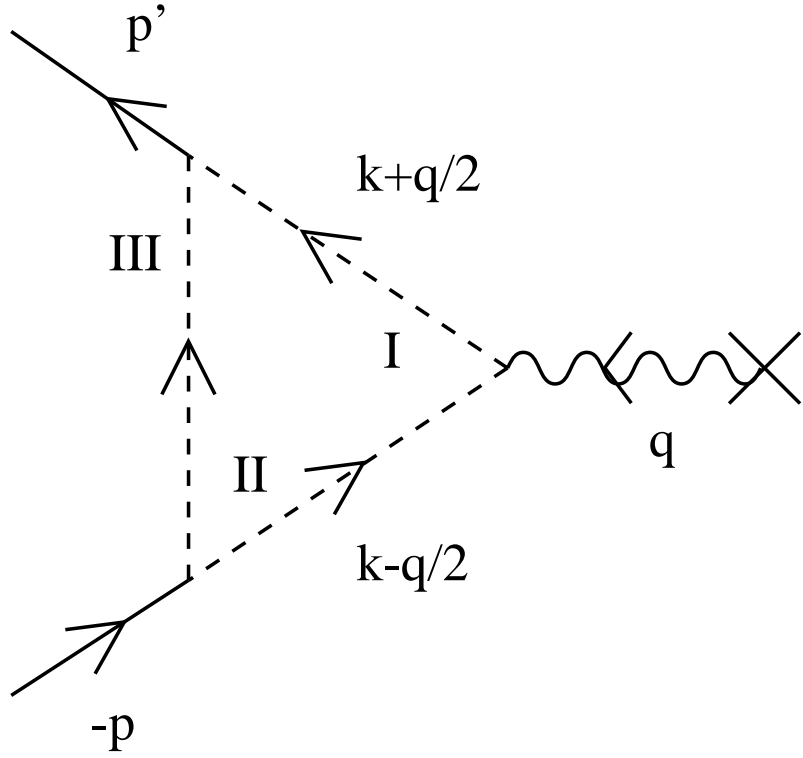}
\end{center}
\caption{\label{fig-2} General form of the calculated loop diagrams.
The various combinations of intermediate particles considered in
our calculation are listed in Table \protect\ref{kstartab1}.}
\end{figure}

\end{document}